\documentclass[structabstract]{aa}  
\usepackage{graphicx}
\usepackage{txfonts}
\begin{document}
    \title{Search for radial velocity variations in eight M-dwarfs with NIRSPEC/Keck~II}

\author{F. Rodler\inst{1,2,3}
\and
R. Deshpande\inst{4,5}
\and
M. R. Zapatero Osorio\inst{6}
\and
E. L. Mart\'in\inst{6}
\and
M.M.~Montgomery\inst{7}
\and
C. del Burgo\inst{8} 
\and
O.~L.~Creevey\inst{2,3}}
\offprints{frodler@iac.es}
 \institute{Institut de Ciencies de l'Espai (IEEC-CSIC), Campus UAB, Torre C5 - parell - 2$^a$ planta, 08193 Bellaterra, Spain
    \and
    Instituto de Astrof\'{i}sica de Canarias,
             C/V\'{i}a L\'{a}ctea s/n, E-38205 La Laguna, Spain
\and 
Universidad de La Laguna, Avda. Astrof\'isico
Francisco S\'anchez s/n, 38206 La Laguna, Tenerife, Spain.
     \and
	Center for Exoplanets and Habitable Worlds, The Pennsylvania State University, University Park, PA 16802
\and
Department of Astronomy and Astrophysics, The Pennsylvania State University, University Park, PA 16802
\and
		Centro de Astrobiolog\'ia (CSIC-INTA), Ctra. Ajalvir km. 4, E-28850
		Torrej\'on de Ardoz, Madrid, Spain
\and
University of Central Florida, Physics Department, PO Box 162385, Orlando, FL 32816, USA
\and
UNINOVA-CA3, Campus da Caparica, 2825-149 Caparica, Portugal}

  \date{Received ?; accepted ?}
 
 
  \abstract
   {Radial velocity (RV) measurements from near-infrared spectra have become a potentially powerful tool to search for planets around cool stars and sub-stellar objects. As part of a large survey to characterize M-dwarfs using NIRSPEC at Keck~II, we obtained spectra of eight late M-dwarfs (spectral types M5.0-M8.0) during two or more observing epochs per target. These spectra were taken with intermediate spectral resolving powers ($R\sim20,000$)  in the $J$-band.  }
   {We search for relative radial velocity variability in these late
M-dwarfs and test the NIRSPEC capability of detecting short period brown dwarf
and massive planetary companions around low-mass stars 
in the $J$-band ($\approx 1.25~\mu$m).  Additionally, we
reanalyzed the data of the M8-type star vB10 (one of our targets)
presented in Zapatero Osorio et al. (2009), which were obtained with the
same instrumentation as our data.}
   {To achieve a precise RV measurement stability, the NIRSPEC spectra are self-calibrated by making use of the telluric absorption lines, which are present in the observed spectra and used as a long-term stable reference. In the modeling process a multi-parameter $\chi^2$-optimization is employed to generate an accurate description of the observation. The telluric lines allow us to model the instrumental profile of the spectrograph and the determination of the Doppler shift of the stellar absorption lines. }
   {For the entire M-dwarf sample, we do not find any evidence of relative RV variations induced by a short period brown dwarf or massive planetary companion. The typical RV precision of the measurements is between 180 and 300~m~s$^{-1}$, which is sufficient to detect hot Neptunes around M-dwarfs. Also, we find that the spurious RV shift in Zapatero et al.~(2009) of the star VB10 was caused by asymmetries in the instrumental profile between different observing epochs, which were not taken into account in their analysis.}
   {}

 
   \maketitle
%

\section{Introduction}
The search for extrasolar planets  has
led to more than 700 confirmed discoveries\footnote{The Extrasolar Planets Encyclopedia;
http://www.exoplanet.eu; 2011-Nov-15} by using all detection techniques. Up to now, most of them have been detected by means of the radial velocity (RV) technique using high-resolution spectrographs ($R=\lambda / \Delta \lambda \ge 40,000$)
at optical wavelengths. Most discoveries are giant gaseous planets (typically hot Neptunes and Jupiters) of short periods (of a few days) around stars of spectral 
types F, G and K.

As potential hosts to rocky planetary companions, M-dwarfs have become increasingly popular as
targets for RV searches (e.g. Endl et al. 2006, Charbonneau et al. 2009, Mayor et al. 2009, Zechmeister et al.~2009). Very cool stars such as M-dwarfs are the most abundant type ($\sim70 \%$) of stars in the solar neighborhood and the Milky Way in
general (Henry at al.~1997).
The effective temperatures and masses of M-dwarfs, respectively, are in the range 3700 to 2200~K and 0.5 to 0.07 solar masses for the M0 to M9.5 spectral types. They exhibit prominent absorption features corresponding to strong neutral atoms, H$_2$0, FeH, VO, CO, and TiO.
Owing to the low masses of these
objects, the reflex motion of the host star due to the gravitational pull of the
extrasolar planet is higher and more easily detectable than for more massive
host stars. Since M-dwarfs are very cool stars in comparison with solar-type stars, short period planets would more
likely be situated in the habitable zone.

M-dwarfs emit most of their energy around $1.1-1.3~\mu {\rm m}$, in the near-infrared
(NIR), while they appear very faint at optical wavelengths.
First attempts to measure RV variations among very cool M-dwarfs at NIR wavelengths were done by Mart\'in et al.~(2006). They achieved a RV precision of around 300 m~s$^{-1}$ for the M9.5-dwarf LP944-20 by using the spectrograph NIRSPEC, which is mounted on the Keck II telescope in Hawaii
(McLean et al.~1998).  Recently, several research groups have reported high-precision RV measurements taken in the NIR with CRIRES
(K\"aufl et al.~2004), mounted at the UT1/VLT in the Paranal
Observatory of ESO in Chile. Bean et al.~(2010a) conducted high-resolution spectroscopic data of 
over 60 M-dwarfs (spectral types M4-M9) and used a NH$_3$ gas cell spectrum as a stable reference, and report an
RV precision of better than $5~{\rm m~s^{-1}}$. Figueira et al.~(2010) took observations of the planetary
candidate TW~Hya and achieved a RV precision better than $10~{\rm m~s^{-1}}$ by adopting telluric lines as
a stable reference. Blake et al.~(2010) report RV measurements of 59 M- and L-dwarfs using the Keck/NIRSPEC spectrograph, with the aim to detect low-mass companions. They made use of strong CO absorption features around 2.3~$\mu$m in M- and L-dwarfs and achieved  RV precisions between 50 and 200~m~s$^{-1}$. Tanner et al.~(2010) report preliminary results of a late M-dwarf survey by using Keck/NIRSPEC with RV precisions between 150-300~m~s$^{-1}$.

In 2009, Pravdo \& Shaklan~(2009) announced a massive planet around the M-dwarf vB10 discovered by means of astrometrical data. 
Zapatero Osorio et al.~(2009; hereafter ZO09) made use of our NIRSPEC data set and found evidence for RV variations, which supported the planet hypothesis. They achieved a RV precision of about 300~m~s$^{-1}$. However, this planet was later refuted by different groups: Bean et al. (2010b), who took high-resolution spectra ($R=\lambda/\Delta\lambda\sim100,000$) with CRIRES and who achieved a RV precision of $\sim 10$~m~s$^{-1}$, and by Anglada-Escud\'e et al. (2010). Additionally, Lazorenko et al.~(2011) carried out an astrometric survey using the FORS2 camera of the ESO/VLT on Cerro Paranal, Chile, but found no evidence for the existence of a massive planet orbiting vB10. As part of this work, we aimed at finding out what had caused the spurious RV variations in the data analysis of ZO09. 

Here, we report relative RV measurements of 8 late M-dwarfs with NIRSPEC, and we support the capability of this instrument to detect giant planetary companions with short orbital periods. In Section 2 we describe our M-dwarf sample, our observations and data reduction. In
Section 3 we outline the details of the data analysis, followed by the results and discussion (Section~4).

\section{Observations and Data Reduction}

As part of our M-dwarf survey (Deshpande et al., in prep.), we observed 8 M-dwarfs (2M2331, GJ1156, GJ406, GJ905, LHS1363, RXJ2208.2, and vB10) at two or more epochs (Table~\ref{xoxo:T1}) using the NIRSPEC instrument, mounted on the Keck II telescope on the summit of Mauna Kea in Hawaii (McLean et al. 1998).  We aimed at conducting RV precision tests, and searching for RV drifts which could be interpreted as massive planets orbiting those M-dwarfs. Our sample comprised dwarfs with spectral types of M5.0-M8.0 and masses between 0.14 and 0.075~M$_\odot$. Table~\ref{xoxo:T4} provides a list of the spectral types, $J$-band magnitude and the projected stellar rotational broadening $v \sin i$ of the targets.


NIRSPEC is a cross-dispersed, cryogenic echelle spectrometer employing a $1024\times
1024$ ALADDIN InSb array detector. In the echelle mode, we selected
the NIRSPEC-3 ($J$-band) filter and an entrance slit width of $0.432\arcsec$ (i.e. 3 pixels along the dispersion direction of the detector), except for the 2001 June observations of vB10, where we used an entrance slit width of $0.576\arcsec$.
The corresponding spectral resolving powers were $R= \lambda / \Delta \lambda \approx 22,700$ and $R \approx 17,800$, respectively for the $0.432\arcsec$  slit and the $0.576\arcsec$ slit. The length of
both slits was $12\arcsec$. All observations were carried out at an echelle angle of 
$\sim 63^\circ$. This instrumental setup provided a wavelength coverage from 1.148 to $1.346~\mu$m split into 10 different echelle orders, a nominal dispersion ranging from 0.164 (blue) to 
0.191$~{\rm \AA~pix^{-1}}$ (red wavelengths). Weather conditions (seeing and atmospheric transparency) were fine during the observations, except for the 2008 epoch, which was hampered by cirrus and strong wind. Table~\ref{xoxo:T1} lists the individual exposure times and the  signal-to-noise ratios (SNRs) on average per spectral pixel in the stellar continua for each observing epoch. 

For each target, the spectra were collected at two different positions along the entrance slit. This nodding allowed later the removal of the OH sky emission lines. For the identification of atmospheric telluric absorption, near-infrared featureless stars of spectral types A0-A2 were observed close in time (on average 3 min before or after the target observations) and position to our targets.

    Raw data were reduced using the echelle package within
{\tt IRAF}\footnote{{\tt IRAF} is distributed by the National Optical Astronomy Observatory (NOAO), which is operated by the Association of Universities for Research in Astronomy, Inc., under contract with the National Science Foundation in the USA.}.   Nodded images were
subtracted to remove sky background and dark current. White
light spectra obtained with the same instrumental configuration
and for each target observation were used to flat-field the data.
By adopting the {\tt apall} task, we first identified and optimally centered the echelle orders in the two individual nodding frames for each target and traced these orders by adopting a second-order Legendre polynomial along the dispersion axis. In the next step, we extracted the one-dimensional spectra for each echelle order adopting the same aperture/trace parameters for both, the target as well as an arc-lamp exposure of Ar, Kr, and Xe, which was always acquired after observing
the target and before pointing the telescope at the next star. The
air wavelengths of the arc lines were identified using the NIST\footnote{ http://physics.nist.gov/PhysRefData/ASD/lines\_form.html}
database, and we produced preliminary wavelength calibration fits using a third-order Legendre
polynomial along the dispersion axis and a second-order one
perpendicular to it. The mean rms of the fits was $0.03~{\rm \AA}$, or
0.7 km~s$^{-1}$. 


\begin{table}
\begin{center}
\caption{Journal of target observations. The average SNR values in the stellar continua are given.} \label{xoxo:T1}
\begin{tabular}{lcccc}
\hline\hline
Obs. date & UT & Slit & Exp. (s) & SNR$^{a}$\\
\hline
\multicolumn{2}{l}{{\bf 2MJ2331-2749}:} \\
2007-Jun-24 & 14:33 &  $0.432 \times 12$ &  $2\times200$ & $\sim50$ \\
2007-Jun-25 & 14:12 &  $0.432 \times 12$&  $2\times200$ & $\sim60$ \\
\hline
\multicolumn{2}{l}{{\bf GJ 406}:} \\
2007-Apr-30 & ~7:03 &  $0.432 \times 12$ & $2\times120$ & $\sim50$  \\
2007-Dec-23 & 15:32 &  $0.432 \times 12$ & $2\times~30$ & $\sim50$  \\
\hline
\multicolumn{2}{l}{{\bf GJ 905}:} \\
2007-Jun-25 & 14:42 &  $0.432 \times 12$ & $2\times~20$  & $\sim110$ \\
2007-Oct-27 & 10:51 &  $0.432 \times 12$ & $2\times120$  & $\sim280$ \\
\hline
\multicolumn{2}{l}{{\bf GJ 1156}:} \\
2007-Jun-24 & ~7:24 & $0.432 \times 12$  & $2\times120$ & $\sim140$  \\
2007-Jun-25 & 15:55&  $0.432 \times 12$ & $4\times300$ &  $\sim240$  \\
\hline
\multicolumn{2}{l}{{\bf LHS 1363}:} \\
2007-Oct-26 & 12:09 &  $0.432 \times 12$ & $2\times300$  & $\sim110$ \\
2007-Oct-27 & 12:13 &  $0.432 \times 12$ & $2\times300$  & $\sim110$ \\
\hline
\multicolumn{2}{l}{{\bf LP 412-31}:} \\
2007-Oct-26 & 12:44 & $0.432 \times 12$  & $2\times300$  & $\sim60$ \\
2007-Oct-27 & 12:56 & $0.432 \times 12$  & $2\times300$  & $\sim50$ \\
\hline
\multicolumn{2}{l}{{\bf RXJ2208.2}:} \\
2007-Jun-24 & 13:49 &  $0.432 \times 12$ & $2\times120$  & $\sim70$ \\
2007-Jun-25 & 13:40 &  $0.432 \times 12$ & $2\times120$  & $\sim70$ \\
\hline
\multicolumn{2}{l}{{\bf vB10}:} \\
2001-Jun-15 & 14:06 &  $0.576 \times 12$ & $2\times100$  & $\sim60$ \\
2001-Nov-~2 & ~4:43 &  $0.432 \times 12$ & $2\times120$  & $\sim60$ \\
2001-Nov-~2 & ~5:39 &  $0.432 \times 12$ & $2\times120$  & $\sim60$ \\
2007-Jun-25 & 13:22 &  $0.432 \times 12$ & $2\times120$  & $\sim70$ \\
2008-Jul-28 & ~6:07 &  $0.432 \times 12$ & $2\times120$  & $\sim20$ \\
\hline
\hline
\end{tabular}
~~~~~~~~~~~~~~~~~$^{a}$ SNR on average in the pseudo stellar continuum per spectral pixel around 1.265~$\mu$m.\\
\end{center}
\end{table} 

\begin{table}
\begin{center}
\caption{Properties of the M-dwarfs} \label{xoxo:T4}
\begin{tabular}{lcccccl}
\hline
\hline
Name & Sp. type & $J$ & $v \sin i$ & Ref. \\
& & & (km s$^{-1}$)  \\
\hline
2MJ2331-2749 & M7.0 & 11.65 & $<12$ & Des11 \\
GJ406 & M5.5 & 7.09 & $\sim3$ &Rei10 \\
GJ905 & M5.0 & 6.88 & $<3$ & Rei10 \\
GJ1156 & M5.0 & 8.52 & $17.2\pm2.9$ & Des11 \\
LHS1363 & M6.5 & 10.48 & $<12$ & Des11 \\
LP412-31 & M6.5 & 10.48 & $17.6\pm3.2$ & Des11 \\
RXJ2208.2 & M5.0 & 10.60 & $18.6\pm2.3$ & Des11 \\
vB10 & M8.0 & 9.91 & 6.5 & Moh03 \\
\hline
\hline
\multicolumn{5}{l}{Abbreviations:}\\
\multicolumn{5}{l}{Des11 ...... Deshpande et al., in prep.}\\
\multicolumn{5}{l}{Moh03 ...... Mohanty \& Basry (2003).}\\
\multicolumn{5}{l}{Rei10 ...... Reiners et al. (2010).}\\
\end{tabular}
\end{center}
\end{table}

\section{Relative Radial Velocity Method}
We measured the RV of the stars relatively to the telluric lines present in the 
spectra as well as to a selected epoch of the star, employing a self-calibration approach.
 The radial velocity of
the telluric lines is constant in all wavelengths down to a level
of 10~m~s$^{-1}$ (e.g., Figueira et al. 2010, Seifahrt \& K\"aufl 2008), which is about a magnitude 
smaller than the velocity precision we can achieve with NIRSPEC. 
The basics of the self-calibration method have been extensively described 
(e.g. Valenti et al.~1995, Endl et al. 2000, Bean et al.~2010a), so that we just 
give a concise description of the method here and point out the important 
aspects of its implementation.

Briefly, the main idea is to model the observations 
and thereby determine the  relative RV shift, and perform a fine tuning 
of the wavelength solution at the same time. Basically, the model spectrum 
is the product of a high-resolution telluric spectrum with a Doppler-shifted 
version of a high-resolution reference spectrum of the star. This product of 
those two spectra is then subjected to a convolution with the instrumental 
profile (IP) of the spectrograph, and finally binned to the sampling of the 
observed data. By variation of the free parameters of the model (Table~\ref{xoxo:T2}), 
the best fit model is evaluated by $\chi^2$ statistics. 
The input Doppler-shift which yields the best fit represents the measured RV. 

Since our method requires the presence of telluric lines in the spectra, we restricted the analysis to the echelle orders 66, 60, 58, and 57, which were heavily contaminated mainly by absorption lines of water vapor. These four orders correspond to the wavelength ranges of $\lambda \sim 1.147$ to 1.163~$\mu$m, 1.261 to 1.279~$\mu$m, 1.304 to  1.323~$\mu$m, and  1.327 to  1.346~$\mu$m, respectively. For the echelle order numbering we refer to McLean et al.~(2007). Subsequently, we subdivided each spectral order into 5 equidistant pixel chunks of 200 pixel each (i.e. for all four orders together we have 20 chunks). This step was done to simplify the process of improving the model, to speed up the calculations and to account for variations of the IP throughout each spectral order. Each of the following steps was carried out on each chunk individually, and the SNR was determined by
\begin{equation} \label{xoxoequ:50}
{\rm SNR} = S_{\star} / \sqrt{S_{\star} + S_{\rm BG} + {S_{\rm BG2} + \rm RON}^2\times 2n} ,
\end{equation}
where $S_{\star}$ denotes the signal level from a star in electrons, integrated over an aperture of $n$ pixels, 
$S_{\rm BG}$ is the signal level of the sky background, $S_{\rm BG2}$ is the signal level from the sky background of the frame taken at the other nodding position, and RON denotes the read-out-noise level per pixel in rms electrons (for NIRSPEC, RON=65e$^{-1}$). The noise errors were propagated in the following data analysis steps.

\subsection{Step 1: Telluric template spectrum and determination of the instrumental profile}
For the calculation of the atmospheric transmission spectrum, we used the Line-By-Line Radiative Transfer Model (LBLRTM) code, which is based on the FASCODE algorithm (Clough et al.~1992). LBLRTM is available as fortran source code \footnote{Source code and manuals are available under {\tt http://rtweb.aer.com/lblrtm\_description.html}} and runs on various platforms. As molecular database we adopted HITRAN (Rothman et al.~2005), which contains the 42 most prominent molecules and isotopes present in the atmosphere of the Earth. Following the approach presented by Seifahrt et al.~(2010), we created a high-resolution theoretical telluric spectrum for each observed spectrum by accounting for the air mass of the star as well as the weather conditions (water vapour density column, temperature and pressure profiles) during the observations. We retrieved the weather information from the Global Data Assimilation System (GDAS). GDAS models are available in 3 hours intervals for any location around the globe\footnote{GDAS webpage: {\tt http://ready.arl.noaa.gov/READYamet.php}}.

\begin{figure}
\includegraphics[angle=0,scale=.45]{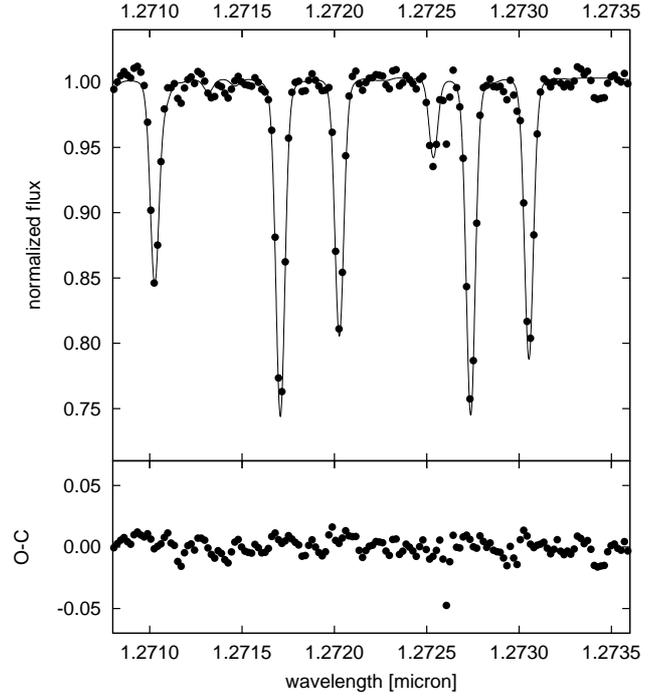}
\caption{Comparison between the observed telluric spectrum (points), and the theoretical model (line). The observed telluric spectrum was taken by using a featureless A-star (HD181414), which was observed with NIRSPEC in the $J$-band on 2007-06-25. The rms of the telluric model fit to the observed data is about 1\%. \label{xoxo:F1A}}
\end{figure}

To calculate a first version of the instrumental profile (IP) of the spectrograph, we made use of the A-star observations next to our targets. First of all, we normalized the spectrum in such a way that the flux in the telluric continuum was at one. Next, we refined the wavelength solution of the observed telluric spectrum with the appropriate high-resolution theoretical telluric spectrum by adopting a second order polynomial. We then determined a preliminary version of the IP as the sum of 7 Gaussian profiles in a similar way as described in Valenti et al.~(1995): Around a central Gaussian we grouped 3 Gaussians on each side of it, which allow to account for asymmetries in the IP. Free parameters were the height and width of the central Gaussian, plus the heights of the six satellite Gaussians (c.f. Table~\ref{xoxo:T2}). To reduce the number of free parameters per chunk, and to ensure that the method works robust, the positions and the widths of these satellites were fixed and set {\it a priori} in such a way that their half-widths overlapped.

Next, we convolved the high-resolution theoretical spectrum with the determined preliminary IP and compared the resulting spectrum with the observed A-star spectrum. For a few telluric lines, we realized that the theoretical spectrum under- or overestimated the line-depths. To produce a better match between theory and observation, we iteratively carried out a fine tuning of the line-depths in the high-resolution theoretical telluric spectrum, then again convolved the modified telluric spectrum with the IP and evaluated the result with the observation by means of $\chi^2$-statistics. The iterations were carried out until the reduced $\chi^2$ reached 1. Fig.~\ref{xoxo:F1A} shows a comparison between an A-star spectrum (HD~181414) and the fit of the refined theoretical telluric spectrum as well as the residuals of the model fit to the A-star spectrum. 
The rms of the telluric model fit to the observed data is about 1\% on average. The refined high-resolution telluric model spectrum from now on served the purpose of the telluric template spectrum. In the final step, we refined the wavelength solution of the observed spectrum, and then re-calculated the IP by adopting this new telluric template spectrum.

\subsection{Step 2: Stellar template spectrum}

Due to lack of appropriate theoretical model spectra which fit the stellar absorption features in the $J$-band, we created the stellar template spectrum for one selected reference epoch by calculating an IP-free and telluric-free version of the target spectrum. Concerning the reference epoch, we selected that epoch in which the stellar spectrum showed the highest SNR. To produce the stellar template, we first applied the refined wavelength solution of the A-star spectrum (which was taken - on average - 3 minutes before or after the target observations) to the observed target spectrum of the same epoch. Since the telluric lines were present in the target spectrum, we needed to remove them from the spectrum. In preparation for this, we convolved the appropriate theoretical telluric spectrum with the IP. Then, we divided the target spectrum by the convolved theoretical telluric spectrum (Fig.~\ref{xoxo:F1A}). Similar to Bean et al.~(2010a) and Blake et al.~(2010), we found that this approach led to smaller uncertainties than when the usual method of the telluric lines removal was carried out, where the target spectrum is simply divided by the appropriate A-star spectrum.

To create the final stellar template spectrum, we deconvolved the telluric-free target observation by the IP by employing the maximum-entropy method (MEM) with 5 times oversampling of the output spectrum. In the final step, we applied the refined wavelength solution that we had obtained for the A-star spectrum to the 5-times over-sampled IP-free stellar spectrum, which from there served the purpose of the stellar template.

\begin{figure}
\includegraphics[angle=0,scale=.5]{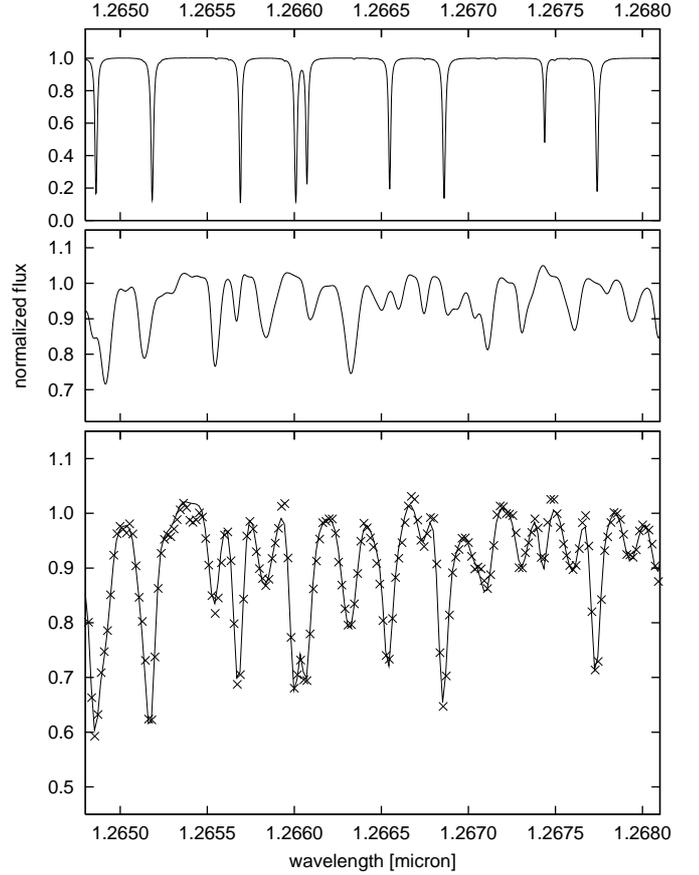}
\caption{Example model components and fit for the radial velocity measurements. The components are given in the two top panels: the spectrum of the high-resolution theoretical telluric spectrum (top), and the deconvolved and RV-shifted  version of the telluric free stellar spectrum (bottom). We note that the scale of the flux is different in each panel for better visibility. In the lower panel, we show the observed spectrum (points) and the best-fit model (line).\label{xoxo:F3A}}
\end{figure}

\begin{table}
\begin{center}
\caption{Free parameters in the model per chunk.} \label{xoxo:T2}
\begin{tabular}{lc}
\hline\hline
Parameter & degree of \\
 ~ & freedom\\
\hline
Stellar absorption line depth & 1 \\
Linear stellar continuum trend & 2 \\
Doppler shift of stellar template & 1 \\
Telluric absorption line depth & 1 \\
Amplitude and width of main Gaussian & 2 \\
Amplitudes of satellite Gaussians & 6 \\
${\rm2^{nd}}$ order wavelength solution vector & 3\\
\hline
\hline
\end{tabular}
\end{center}
\end{table}

\subsection{Step 3: Fitting the observed data}

For each target, we first determined the barycentric velocity differences $\Delta v_{{\rm bc},t}$ for all observation epochs $t$ with respect to that one of the stellar template epoch. This correction was calculated by use of the JPL ephemeris DE200 (Standish 1990). 

We constructed the model of the observation by multiplying the telluric template with the stellar template, where its Doppler shift is one of the free parameters.  
The resulting combination spectrum was subjected to a convolution with the IP that was determined in Step~1, and a new wavelength solution was calculated. Subsequently, all the free parameters (i.e. line-depths in the models, IP, ...; See Table~\ref{xoxo:T2})  were refined by employing Brent's optimization algorithm, and the fit to the observed data was evaluated by using $\chi^2$ statistics. The search range for the Doppler shift was $\Delta v_{{\rm bc},t}\pm15$~km~s$^{-1}$ with a step width of 10~m~s$^{-1}$. We note that such a large interval would also allow us to detect large
relative RV variations due to unseen massive companions such as low-mass
stars and brown dwarfs. We calculated the $\chi^2$ values for each Doppler shift and then determined the exact $\chi^2$-minimum by using a Gaussian fit. That Doppler shift which led to the overall best fit model  ($\chi^2$-minimum) constituted the measured RV of the star in the chunk, relatively to the stellar template.

To determine the global (i.e. all chunks together) RV measurement, we combined 
all the RV measurements in all chunks into one by considering the following restrictions: 
Each chunks was given a specific weight which was determined from the average SNR 
in the stellar continuum, plus the number of telluric lines and stellar absorption lines which 
were present in that chunk, plus the depths of the stellar lines. Furthermore, we rejected chunks in which the RV-measurement constituted clearly an
outlier ($3 \sigma$ above / below average of all RV measurements) by adopting sigma-clipping. No chunks were rejected for the stars GJ905, GJ1156, LHS1363, RXJ2208.2, and vB10. For 2MJ2331-2749 and LP412-31, one chunk was rejected each, while for GJ406 two chunks were rejected. All those rejected chunks were located in noisy areas with SNR levels lower than 40 on average. We attribute these spurious RV shifts to improper stellar templates which contained artifacts coming from the deconvolution of low-SNR data.
 
The global RV measurement 
was then determined as the arithmetic weighted mean of the un-rejected chunks. 
The error of the global RV measurement was determined as the weighted standard deviation 
of the un-rejected RV measurements in the chunks.

\section{Results and Discussion}

We analyzed the data sets with our relative radial velocity measurement approach and determined the relative RV measurements with respect to the selected reference epoch. For any of the eight M-dwarfs in our sample, we have not found significant
evidence of relative RV variations at the level of 3$\sigma$ (Table~\ref{xoxo:T3}), where
$\sigma$ stands for the measurements uncertainty. The RV precisions are in the order of 180-300~m~s$^{-1}$, except for the observations in July 2008, which were taken at low SNR.

\begin{figure}
\includegraphics[angle=270,scale=.4]{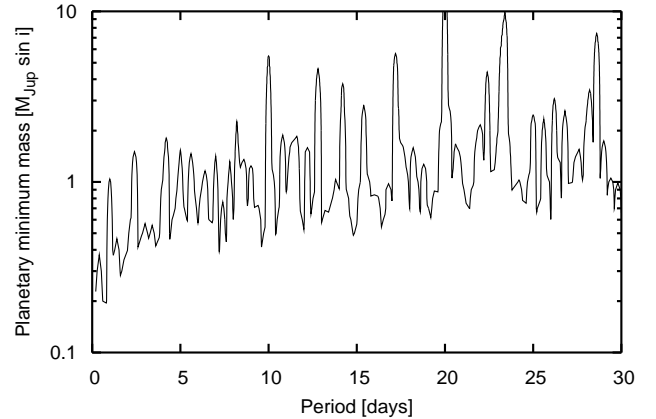}
\caption{Monte-Carlo analysis for the five vB10 measurements. For the mass of vB10, we adopted 0.078~M$_\odot$ from Pravdo \& Shaklan (2009). It is shown that with a RV precision of $\sim220$~m~s$^{-1}$ even hot Jupiters with minimum masses $m_{\rm p} \sin i> 0.3~$M$_{\rm Jup}$ could be detected around late M-dwarfs with 3$\sigma$ confidence. We note that for a larger number of measurements the number of aliasing peaks can be significantly decreased.\label{xoxo:F1}}
\end{figure}

We investigated the period and mass range of companions which could be detected with such RV precisions. We determined the minimum mass of the planet by employing a Monte-Carlo analysis, thereby probing planetary orbits with different parameters and investigating how many of these orbits could be recovered for the five measurements of vB10. We considered only the case of a circular orbit and the mass of vB10, which is $m_\star=0.078~{\rm M}_\odot$.
Fig.~\ref{xoxo:F1} shows the 3$\sigma$ detection limit.  We find that for companions with only a few days period, even planets with minimum masses of $m_{\rm{p}}\sin i \ge 0.3~M_{\rm Jup}$ can be detected with a RV precision of $\sim220$~m~s$^{-1}$.

In Fig.~4 we show our relative RVs of vB10 and the measurements by ZO09.
We note that for a proper comparison, we adopted the same reference epoch
as in ZO09. The agreement between ZO09 and our measurements is within
1$\sigma$ of the quoted uncertainties for all epochs except for the 2001 epoch
(BJD = 2452076). We provide next an explanation for the discrepancy of this
one measurement. Similar to our data analysis, ZO09 used the telluric lines present in the target spectra as a stable reference, but contrary to our analysis they did not account for any IP variations in their analysis, but calculated the RVs by cross correlation. In our analysis, we do not see any RV shift exceeding the RV-precision for any measurement. We get evidence that the different instrumental setting used on 2001-Jun-15 (0.576" slit instead of the standard setting of 0.432")  produced an asymmetric instrumental profile (Fig.~\ref{xoxo:F2}), which led to a significant RV shift when a simple cross correlation is adopted for the RV determination. Our results clearly demonstrate the importance of modeling the IP especially when observations are carried out with different instrumental settings.

\begin{figure}
\includegraphics[angle=270,scale=.355]{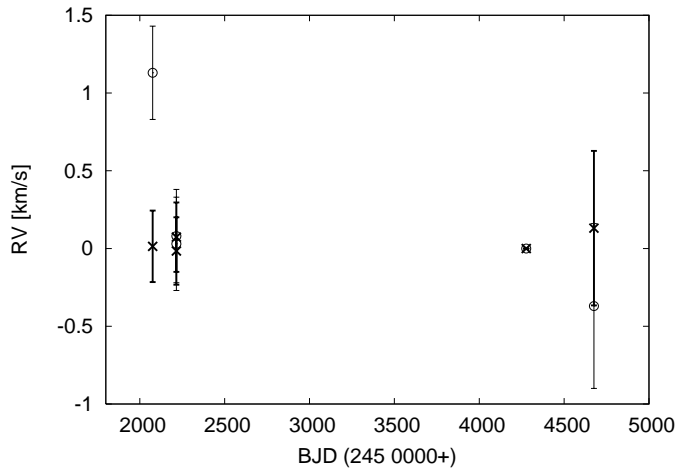}
\caption{Our RV measurements of vB10 (crosses) vs. the measurements of ZO09 (open circles). The RV precision given by our analysis is about $~220$~m~s$^{-1}$, except for the last epoch in 2008, which was hampered by bad weather. We prove that the RV shift in the work of ZO09 in the first 2001 epoch (BJD$=245~2076$) originates from
unaccounted asymmetries in the IP rather than from a planetary companion. 
 \label{xoxo:F2A}}
\end{figure}

\begin{figure}
\includegraphics[angle=270,scale=.36]{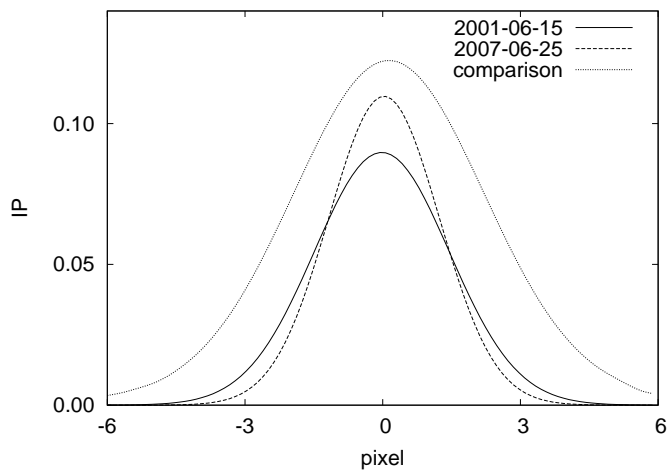}
\caption{Instrumental profiles (IPs) of NIRSPEC for two different observing epochs of vB10. On 2001-06-15, a broader slit was used ($0.576"$; solid line) than for the reference epoch ($0.432"$; 3007-06-25; dashed line). To visualize the asymmetries between both IPs, we calculated the ratio between both IPs and scaled the resulting function for better visibility (dotted line). ZO09 did not account for these asymmetries between both IPs, which led to a spurious RV shift of about 1~km~s$^{-1}$ for the 2001-06-15 measurement in their data analysis.\label{xoxo:F2}}
\end{figure}

\begin{table}
\begin{center}
\caption{Relative radial velocity measurements.} \label{xoxo:T3}
\begin{tabular}{lcccc}
\hline\hline
Obs. Date & BJD & rel. RV \\
 & 245~0000+ & (m~s$^{-1}$) \\
\hline
\multicolumn{2}{l}{{\bf 2MJ2331-2749}:} \\
2007-Jun-24 & 4276.10874 & $194\pm224$\\
2007-Jun-25 & 4277.09424 & {\it ref. epoch}\\
\hline
\multicolumn{2}{l}{{\bf GJ 406}:} \\
2007-Apr-30 & 4220.79751 & {\it reference epoch} \\
2007-Dec-23 & 4458.14962 & $-77\pm 238$ \\
\hline
\multicolumn{2}{l}{{\bf GJ 905}:} \\
2007-Jun-25 & 4277.11249 & $-233\pm 201$\\
2007-Oct-27 & 4400.95657 & {\it reference epoch} \\
\hline
\multicolumn{2}{l}{{\bf GJ 1156}:} \\
2007-Apr-30 & 4220.81513 & $141\pm185$\\
2007-Dec-22 & 4457.16844 & {\it reference epoch} \\
\hline
\multicolumn{2}{l}{{\bf LHS 1363}:} \\
2007-Oct-26 & 4400.01234 & $-27\pm196$\\
2007-Oct-27 & 4401.01511 & {\it reference epoch}\\
\hline
\multicolumn{2}{l}{{\bf LP 412-31}:} \\
2007-Oct-26 & 4400.03654 & {\it reference epoch} \\
2007-Oct-27 & 4401.04491 & $298\pm260$\\
\hline
\multicolumn{2}{l}{{\bf RXJ2208.2}:} \\
2007-Jun-24 & 4276.07879 & $-189\pm 307$\\
2007-Jun-25 & 4277.07266 & {\it reference epoch}\\
\hline
\multicolumn{2}{l}{{\bf vB10}:} \\
2001-Jun-15 & 2076.08951 & $19\pm230$ \\
2001-Nov-02 & 2215.70560 & $69\pm223$ \\
2001-Nov-02 & 2215.74225 & $-3\pm217$ \\
2007-Jun-25 & 4277.05865 & {\it reference epoch} \\
2008-Jul-28 & 4675.75669 & $131\pm497$ \\
\hline
\end{tabular}
\end{center}
\end{table}

We compare our results to the work of Blake et al.~(2010), who searched for companions to M- and L-dwarf wby using NIRSPEC at a spectral resolving power of $\sim25,000$ in the $K$-band. They adopted one spectral order covering the wavelength range from 2.285 to 2.318~$\mu$m to measure the dense and strong CO-absorption line pattern present in those dwarfs. As a stable wavelength reference, they made use of the CH$_4$ telluric absorption lines present in the observations. Similarly to us, they employed a self calibrating approach, with the difference that they adopted theoretical models for M- and L-dwarfs, which well-described the observations.  

Blake et al. obtained measurements with SNRs in the range of 50 to 100 in the pseudo stellar continua, and they report RV precisions of 100-300 m/s for slowly rotating late-M and L dwarfs. The uncertainty of Blake et al. in the K-band is in agreement 
with our derivation of 180-300 m/s in view of our SNRs. However, our wavelength coverage is about twice that in Blake et al. According to the relative RV precision formulae, we should have obtained better velocity precision in terms of wavelength coverage, which is not the case. We conclude that both the larger number of deep lines (more than 30 lines with a line depth larger than 50\%) in the CO-band region as compared to the J-band (only a few lines with a depth larger than 25\%) as well as the use of theoretical template spectra instead of deconvolved stellar spectra  appear to compensate for the shorter wavelength coverage in a similar factor (c.f. Equation~6 in Butler et al. 1996).

We note that Reiners et al.~(2010) and Rodler et al.~(2011) carried out theoretical RV precision studies of M- and L-dwarfs, by adopting theoretical models of M-dwarfs (e.g. del Burgo et al.~2009). As result, they find that the highest RV precision for M-dwarfs is attained in the $Y$ band around $1~\mu$m, rather than in the $J-$ , $H-$ or $K$-band. For L-dwarfs, however, Rodler et al.~(2011) reported that the highest RV precision is attained in the $J$-band.

We conclude that for an accurate relative RV determination with NIRSPEC, a
self-calibrating approach, which accounts for changes in the instrumental
setting, produces the best measurements in terms of RV precision. Although with our RV precision we would be able to detect massive hot Neptunes around late M-dwarfs, we have not found any brown dwarf or massive planetary companion in our survey. Additionally, the re-analysis of the data of the M8-dwarf vB10 presented in ZO09 now clearly confirms the non-existence of a massive planet orbiting that dwarf and agrees with the results by other research groups (e.g. Anglada-Escud\'e et al.~2010; Bean et al.~2010b; Lazorenko et al.~2011).

  \begin{acknowledgements}
We thank to those of the Hawaiian ancestry on whose sacred mountain we are privileged to be guests.
We are grateful to H. Bouy, N. Dello-Russo, P.-B. Ngoc, R. Tata, and R. Vervack for helping to obtain the 2007 and 2008 NIRSPEC spectra.  
FR thanks to A. Seifahrt for his help with LBLRTM, and to M. Zechmeister and M. Endl for discussions on the self-calibrating approach. This work has been supported by the Spanish Ministerio de Eduaci\'on y Ciencia through grant AYA2007-67458. Partial support for this research was provided by RoPACS,
a Marie Curie Initial Training Network funded by the European
Commission's Seventh Framework Programme. The Center for Exoplanets and Habitable Worlds is supported by the Pennsylvania State University, the Eberly College of Science and the Pennsylvania Space Grant Consortium. This work was partly funded by the Funda\c{c}\~{a}o para
a Ci\^{e}ncia e a Tecnologia (FCT)-Portugal through the 
project PEst-OE/EEI/UI0066/201.
We would furthermore like to thank the anonymous referee for valueable comments to improve the article.

  \end{acknowledgements}





\end{document}